\documentclass[showpacs,amssymb,10pt,aps,prd,longbibliography,reprint,nofootinbib]{revtex4-1}

\usepackage{graphicx,epsfig,amssymb,times} 
\usepackage{amsmath,amsfonts}
\usepackage{bm}
\usepackage{epstopdf}
\usepackage[linktocpage,colorlinks]{hyperref}
\usepackage[caption=false]{subfig}
\usepackage[usenames]{color}

\definecolor{coolblack}{rgb}{0.0, 0.18, 0.39}
\definecolor{darkred}{rgb}{0.5,0,0}
\definecolor{darkgreen}{rgb}{0,0.5,0}
\definecolor{darkblue}{rgb}{0,0,0.5}
\definecolor{lapislazuli}{rgb}{0.15, 0.38, 0.61}
\definecolor{venetianred}{rgb}{0.78, 0.03, 0.08}
\definecolor{bleudefrance}{rgb}{0.19, 0.55, 0.91}
\definecolor{dogwoodrose}{rgb}{0.84, 0.09, 0.41}
\hypersetup{colorlinks=true, citecolor=darkblue, linkcolor=darkblue, 
urlcolor = darkblue}

\def\be{\begin{equation}}
\def\ee{\end{equation}}

\newcommand{\bea}{\begin{eqnarray}}
\newcommand{\eea}{\end{eqnarray}}
\newcommand{\ben}{\begin{enumerate}}
\newcommand{\een}{\end{enumerate}}
\newcommand{\bi}{\begin{itemize}}
\newcommand{\ei}{\end{itemize}}

\newcommand{\nn}{\nonumber}

\newcommand{\rs}{R_{\rm S}}
\newcommand{\mh}{M_{\rm BH}}

\def\ga{\mathrel{\raise.3ex\hbox{$>$\kern-.75em\lower1ex\hbox{$\sim$}}}}
\def\la{\mathrel{\raise.3ex\hbox{$<$\kern-.75em\lower1ex\hbox{$\sim$}}}}

\def\l{\left}
\def\r{\right}
\def\be{\begin{equation}}
\def\ee{\end{equation}}

\def\I_M{{I_{\scriptscriptstyle M\times M}}}

\def\be{\begin{equation}}
\def\ee{\end{equation}}
\def\bea{\begin{eqnarray}}
\def\eea{\end{eqnarray}}
\newcommand{\beq}{\begin{eqnarray}}
\newcommand{\eeq}{\end{eqnarray}}
\def\pa{\partial}

\newcommand{\beqal}{\begin{eqnarray}\label}
\newcommand{\beqa}{\begin{eqnarray}}
\newcommand{\eeqa}{\end{eqnarray}}

\begin{document}
\title{\large Absorption by dirty black holes: Null geodesics and scalar waves}

\author{Caio F. B. Macedo}\email{caiomacedo@ufpa.br}
\author{Luiz C. S. Leite}\email{luiz.leite@icen.ufpa.br}
\author{Lu\'is C. B. Crispino}\email{crispino@ufpa.br}
\affiliation{Faculdade de F\'{\i}sica, Universidade 
Federal do Par\'a, 66075-110, Bel\'em, Par\'a, Brazil.}

\begin{abstract}
Black holes are a paradigm in physics nowadays and are expected to be hosted at the center of galaxies. Supermassive galactic black holes are not isolated, and their surroundings play crucial roles in many observational features. The absorption and scattering of fields by isolated black holes have been vastly studied, allowing the understanding of many phenomenological features. However, as far as we are aware, a study of the influence of the presence of matter surrounding black holes in their planar wave scattering and absorption spectrum is still lacking in the literature. This may be important in the analysis of, for instance, the accretion of dark matter by black holes. We consider planar massless scalar waves incident upon a Schwarzschild black hole surrounded by a thin spherical shell. We use the partial-wave method to determine the absorption cross section and present a selection of numerical results. In the low-frequency regime, we show that the absorption cross section is equal to the horizon area. At the high-frequency regime, we show that the absorption cross section approaches the geodesic capture cross section.

\end{abstract}

\pacs{
04.70.-s, 
04.70.Bw, 
11.80.-m, 
04.30.Nk, 
}
\maketitle

\section{Introduction}

In the year of 2015, the Theory of General Relativity (GR) reaches 100 years of existence. In its centennial history, GR has been submitted to many experimental tests~\cite{willreview,psaltisreview} and has obtained remarkable success in all of them. Among the predictions of GR, black holes (BHs) arise as one of the most curious and fascinating ones, due to the features presented by these objects. Within GR, one can state (under certain conditions) that in electrovacuum, isolated black holes are governed by just three parameters~\cite{lrr-2012-7}: mass, charge, and angular momentum. In astrophysical environments, BHs are likely to be surrounded by a rich structure, e.g., the accretion disks~\cite{Narayan:2005ie}. 

The surroundings of BHs have a major importance in many of their observational features. Some gravitational features considering matter surrounding BHs were studied in the past years. Configurations known as \textit{dirty} BHs were considered to analyze the influence of environmental matter around BHs~\cite{Medved:2003rga}. In Ref.~\cite{Leung} a perturbative formula to study quasinormal modes of BHs with surrounding matter was proposed, where it was shown that the presence of an environment can modify the quasinormal modes of the BHs. More recently, in Ref.~\cite{Barausse}, the BH environment influence in the gravitational wave phenomenology was studied in a broad class of scenarios, some of them in which the whole configuration resembles an isolated BH. Additionally, in Ref. \cite{Gurlebeck} G\"urlebeck has argued that the presence of matter surrounding a Schwarzschild BH has no influence on the multipole moments of the distorted BH, within generic assumptions on the matter.

The absorption and scattering of fields have been widely investigated for isolated Schwarzschild~\cite{Sanchez:1977si, Doran:2005vm, Crispino:2007qw, Crispino:2009xt,Decanini:2011xi}, Reissner-Nordstr\"om~\cite{Jung:2005mr, Crispino:2008zz, Crispino:2009ki, Crispino:2009zza,Oliveira:2011zz, Benone:2014qaa,Benone:2015bst}, Kerr~\cite{Glampedakis:2001cx, Dolan:2008kf, Caio:2013}, and regular BHs~\cite{PhysRevD.90.064001,PhysRevD.92.024012}, which helped to understand many of the BH phenomenological features. Although planar wave scattering seems to be a very peculiar phenomenon, many interesting outcomes may be analyzed through it, like the accretion of dark matter by compact objects \cite{Macedo:2013jja,Macedo:2013qea}. Also, planar wave absorption shares many features with the accretion of a uniform velocity fluid into a BH~\cite{Petrich:1988zz}. Moreover, the scattering of light by BHs may cast a shadow \cite{Huang:2007us}, that should be visible with near-future telescopes.

A review on wave propagation, taking into account the coupling with matter and fields, can be seen in Ref.~\cite {Thorne}. However, the study of the gravitational backreaction of the matter surrounding the BH in its absorption cross section is still lacking in the literature. In this paper, we take a first step in this line of investigation, analizing the case of Schwarzschild BHs surrounded by a thin spherical shell \cite{Shell}, which we shall call dirty black holes (DBHs). More specifically, we analyze the absorption cross section of planar massless scalar waves impinging upon a Schwarzschild BH surrounded by a thin shell of matter. We also consider how the dirtiness of the BH can change the absorption of null geodesics.

The remaining of this paper is organized as follows. In Sec.~\ref{sec:metric}, we review some features of a Schwarzschild BH surrounded by a thin spherical shell. In Sec.~\ref{sec:scalarfield}, we investigate the massless scalar field in the spacetime of interest and the suitable boundary conditions for planar wave scattering. In Sec.~\ref{sec:absorption}, we provide expressions for the absorption cross section of a Schwarzschild BH with a thin spherical shell, valid in the low- and high-frequency regimes, with emphasis to the geodesic analysis. In Sec.~\ref{sec:results}, we show a selection of our numerical results, considering different possibilities for the shell position and for the BH mass. We finalize pointing out some remarks in Sec.~\ref{sec:remarks}. Throughout the paper, we use natural units $(c=G=\hslash=1)$. Latin indices at the beginning of the alphabet~$(a, \,b,\, \dots)$ are related to the four-dimensional spacetime metric.

\section{Black holes with a surrounding thin spherical shell}\label{sec:metric}

Spherically symmetric four-dimensional spacetimes can be described by the following line element,
\be
ds^2=-A(r)dt^2+\frac{1}{B(r)}dr^2+r^2d\Omega^2,
\label{eq:metric}
\ee
where $A$ and $B$ are functions of the radial coordinate $r$ only and $d\Omega^2$ denotes the unit 2-sphere line element. We want to describe the metric functions $A$ and $B$ corresponding to a thin spherical shell constituted of a perfect fluid outside a Schwarzschild BH. In GR, (isotropic) fluid configurations are described by the Tolman-Opennheimer-Volkoff (TOV) equations~\cite{Shapiro:1983du}, i.e.,
\bea
\frac{d\mu}{dr}&=&4\pi r^2\rho(r),\label{eq:dmu}\\
\frac{dA}{dr}&=&2\frac{\mu+4\pi r^3 p}{r^2-2 r \mu}A,\label{eq:dA}\\
\frac{dp}{dr}&=&-\frac{\mu +4\pi r^3 p}{r(r-2\mu)}(p+\rho),\label{eq:dp}
\eea
where $p$ and $\rho$ are the pressure and density of the fluid, respectively, and $\mu$ is the mass function, defined through
\be
B(r)=1-\frac{2\mu}{r}\label{eq:B}.
\ee
The vacuum solutions (i.e.,~$\rho=p=0$) of the TOV equations are given by
\bea
A&=&a_1\l(1-\frac{2a_2}{r}\r)\label{eq:sol_A}\\
\mu&=& a_2,\label{eq:sol_mu}
\eea
where $a_1$ and $a_2$ are constants. From Eq.~\eqref{eq:dmu}, we can see that $a_2$ accounts for the total mass-energy of the configuration, within the radius $r$. Moreover, we note that the vacuum solution can always be brought to its Schwarzschild form, both inside and outside the spherical shell, by absorbing the constant $a_1$ in a time reparametrization~\cite{Israel,Israel2}. However, here we choose to keep the constant $a_1$ for the solution inside the shell.

Let us restrict ourselves to the case of a thin spherical shell surrounding a Schwarzschild BH~\cite{Shell}. Let $\rs$ denote the radial position of the shell. We have that, for $r>\rs$, the metric function can be written in the Schwarzschild form:
\be
A(r)=B(r)=1-\frac{2M}{r},
\ee
where $M$ is the Arnowitt--Deser--Misner (ADM) mass. In the region between the BH event horizon and the spherical shell, we can see from Eqs.~\eqref{eq:B} and \eqref{eq:sol_mu}, that the constant $a_2=\mh$ is the mass of the BH as measured by its horizon surface area. We choose the remaining constant $a_1$ such that $A(r)$ is continuous across the spherical shell. We have that
\be
\alpha\equiv a_1 =\frac{(1-2 M/\rs)}{(1-2 \mh/\rs)}.
\label{eq:al}
\ee
Therefore, the full solution is given by
\bea
A(r)&=&\l\{
\begin{array}{l l}
\alpha(1-2\mh/r),&r<\rs\\
(1-2M/r),&r>\rs
\end{array}\r. ,\label{Afunction}\\
B(r)&=&\l\{
\begin{array}{l l}
(1-2\mh/r),&r<\rs\\
(1-2M/r),&r>\rs
\end{array}\r. \label{Bfunction}.
\eea

Using the metric discontinuities, we can evaluate the surface energy $\Sigma$ and the surface tension $\Theta$ of the spherical shell, which are obtained through the following relations \cite{Visser},
\bea
&\sqrt{B(\rs)}_+ - \sqrt{B(\rs)}_- = -4\pi\rs\Sigma,\\
&\l[\frac{A'(\rs)\sqrt{B(\rs)}}{A(\rs)}\r]_+-
\l[\frac{A'(\rs)\sqrt{B(\rs)}}{A(\rs)}\r]_-=\nn\\
&8\pi(\Sigma-2\Theta),\label{smeq:tensioneq}
\eea
where the subscripts $+$ and $-$ indicate the limit $r\to\rs$ from $r>\rs$ and $r<\rs$, respectively. The prime denotes derivative with respect to the coordinate $r$. Using the metric components~\eqref{Bfunction}, we find
\be
\Sigma=\frac{1}{4\pi\rs}\left(\sqrt{1-\frac{2\mh}{\rs}} - \sqrt{1-\frac{2M}{\rs}}   \right).\label{eq:density}
\ee
Moreover, using Eq.~\eqref{smeq:tensioneq}, the surface tension can be written as
\begin{eqnarray}
&\Theta=\frac{1}{16\pi}\left(8 \pi \Sigma - \left[\frac{A^\prime(\rs)\sqrt{B(\rs)}}{A(\rs)}\right]_+\right. \nonumber\\
&+\left. \left[\frac{A^\prime(\rs)\sqrt{B(\rs)}}{A(\rs)}\right]_{-}\right).\label{surfacetension}
\end{eqnarray}
Inserting Eqs.~\eqref{Afunction} and \eqref{Bfunction} in Eq.~\eqref{surfacetension}, we obtain
\be
\Theta=\frac{1}{8\pi\rs}\left(\frac{1-\frac{\mh}{\rs}}{\sqrt{1-\frac{2\mh}{\rs}}}-\frac{1-\frac{M}{\rs}}{\sqrt{1-\frac{2M}{\rs}}}\right).
\ee

\section{Scalar field}\label{sec:scalarfield}
The massless scalar field $\Phi$ is described by the Klein-Gordon equation, namely,
\be
 \frac{1}{\sqrt{-g}}\pa_a \l(\sqrt{-g}g^{ab}\pa_b\Phi\r)=0.
\label{eq:kleingordon}
\ee
Monochromatic scalar waves in spherically symmetric spacetimes can be decomposed as
\be
\Phi(t,\, r,\,\theta,\,\phi)=\sum_{lm}\frac{U_{\omega l}(r)}{r}e^{-i\omega t}Y_{lm}(\theta,\,\phi), \label{fielddecomposition}
\ee
and, using Eq. \eqref{eq:kleingordon}, we obtain the following radial equation for $U_{\omega l}(r)$:
\begin{eqnarray}
&\sqrt{AB}\left(\sqrt{AB}U_{\omega l}^{\prime}\right)^{\prime} \nonumber\\
&+\l\{\omega^{2}-A\left[\frac{l(l+1)}{r^{2}}+\frac{\left(AB \right)^{\prime}}{2Ar}\right]\r\}U_{\omega l}=0.\label{eq:de2}
\end{eqnarray}

Introducing the Regge-Wheeler coordinate $x$, which can be defined through
\be
x \equiv \int dr \frac{1}{\sqrt{AB}},\label{eq:tortoisecoo}
\ee
we can rewrite Eq.~(\ref{eq:de2}) in the form of an one-dimensional Schr\"odinger-like equation, given by
\be
\left[-\frac{d^{2}}{dx^{2}}+V_{l}-\omega^2\right]U_{\omega l}(x)  =  0,\label{radialde}
\ee
where $V_{l}(r)$ is the effective potential, namely,
\be
V_{l}(r) = A\left[\frac{l(l+1)}{r^{2}}+\frac{\left(AB\right)^{\prime}}{2Ar}\right].\label{eq:effpotential}
\ee

\begin{figure}
\includegraphics[width=\columnwidth]{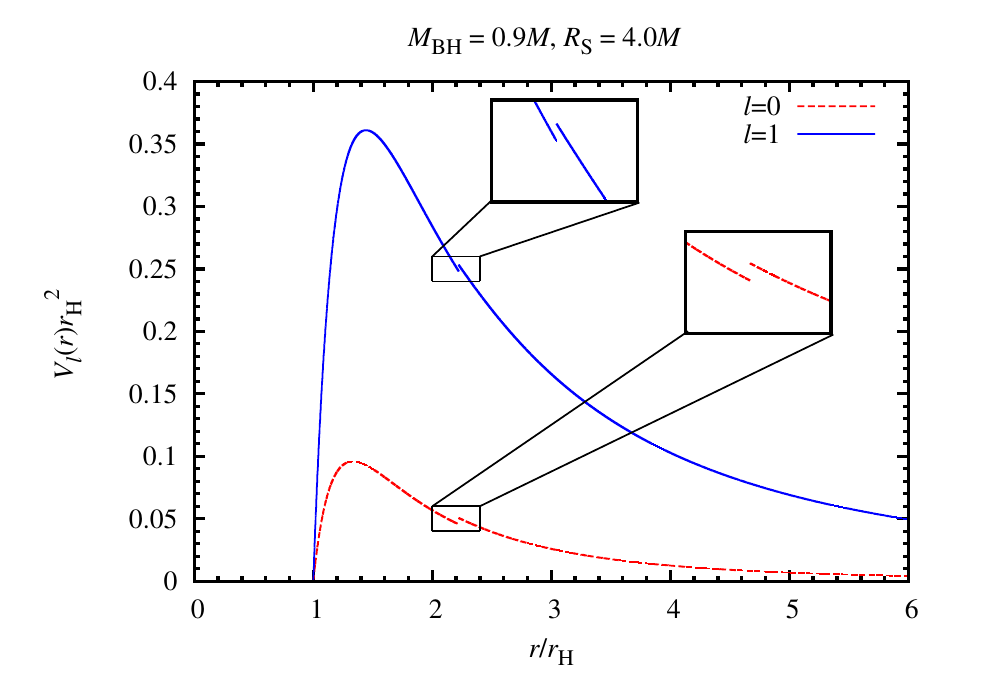}%
\caption{The effective potential, $V_l(r)$, plotted for the modes $l=0,\,1$. Here we have chosen the shell radius $\rs=4.0M$ and the BH mass $\mh=0.9M$. We see clearly the discontinuity of the potential in each case with the help of the insets.}
\label{fig:eff_potential}%
\end{figure}

We note, from FIG.~\ref{fig:eff_potential}, that at the event horizon $r_H=2\mh$ the effective potential vanishes and goes as $1/r^2$ for $r\gg r_H$. These characteristics of the potential are shared with the case of an isolated Schwarzschild BH. However, for DBHs the potential presents a discontinuity at the shell radius, which is visible in~FIG.~\ref{fig:eff_potential}.

The independent solutions of Eq.~(\ref{radialde}) are usually labeled as \textit{in} and \textit{up}. For absorption and scattering, the ones of interest will be the \textit{in} modes, which denote purely ingoing waves impinging from the past null infinity. The in modes obey the following boundary conditions:
\be
U_{\omega l}(x)\sim\l\{
\begin{array}{c l}
	{\mathcal{A}_{\omega l}}R_I+{{\cal R}_{\omega l}} {R_I}^*& (x/M\rightarrow \infty),\\
	{\mathcal{T}_{\omega l}} R_{II} & (x/M\rightarrow -\infty).
\end{array}\r.
\label{inmodes}
\ee
The functions $R_I$ and $R_{II}$ can be written as
\begin{eqnarray}
R_I&=&e^{-i \omega x} {\sum^N_{j=0}} \frac{A^j_\infty}{r^j},
\label{eq:expansion1}\\
R_{II}&=&e^{-i {\omega} x} \sum_{j=0}^N (r-r_{\rm H})^j A^j_{\rm H},
\label{eq:expansion2}
\end{eqnarray}
and the coefficients $A^j_{\infty}$ and $A^j_{\rm H}$ are obtained by requiring the functions $R_I$ and $R_{II}$ to be solutions of the differential equation~(\ref{radialde}), far from the BH and close to the event horizon, respectively. The coefficients ${\mathcal{R}_{\omega l}}$ and ${\mathcal{T}_{\omega l}}$ in Eq.~\eqref{inmodes} are related to the reflection and transmission coefficients, respectively, and obey the following relation:
\be
\l|\frac{{\mathcal{R}_{\omega l}}}{{\mathcal{A}_{\omega l}}}\r|^2=1-\l|\frac{{\mathcal{T}}_{\omega l}}{{\mathcal{A}_{\omega l}}}\r|^2.
\ee

In order to obtain the coefficients ${\cal R}_{\omega l}$ and ${\cal T}_{\omega l}$, we integrate the differential equation from the horizon up to a point far from the configuration (BH with shell). In the integration procedure, one must carefully choose the boundary conditions at the spherical shell. We impose that the scalar field is continuous at the shell location, i.e.,
\be
U_{\omega l}(\rs)_+=U_{\omega l}(\rs)_-.\label{eq:continuity}
\ee
Integrating the differential equation \eqref{eq:de2} across the shell location $r=\rs$, we find the jump condition~\cite{Barausse}:
\begin{align}
&\left[\sqrt{AB}~U_{\omega l}^{\prime}(\rs)\right]_+-\left[\sqrt{AB}~U_{\omega l}^{\prime}(\rs)\right]_{-} \nonumber\\
&=\frac{U_{\omega l}(\rs)}{\rs}\left(\sqrt{AB}_{+}-\sqrt{AB}_{-}\right).\label{eq:jumpcond}
\end{align}
Using the metric functions~(\ref{Afunction}) and (\ref{Bfunction}), the jump condition can be rewritten as
\begin{align}
&\left(1-\frac{2M}{r}\right)U_{\omega l}^{\prime}(\rs)_+-\sqrt{\alpha}\left(1-\frac{2\mh}{r}\right)U_{\omega l}^{\prime}(\rs)_- \nonumber\\  
&=-2\frac{\left(M-\mh\right)U_{\omega l}(\rs)}{\rs^{2}\left(1+1/\sqrt{\alpha}\right)}.\label{eq:jumpcond2}
\end{align}

In practice, one integrates the differential equation $\eqref{eq:de2}$ from very close to $r_H$ up to the spherical shell, using the inner boundary condition, extracting $U_{\omega l}(\rs)_-$ and $U_{\omega l}^{\prime}(\rs)_-$. With these values, we use the jump condition to determine $U_{\omega l}^{\prime}(\rs)_+$ and then integrate again the differential equation \eqref{eq:de2} from the shell to a point far away from the configuration which corresponds to the numerical infinity. The reflection and transmission coefficients can be obtained by comparing the numerical solution with the asymptotic forms \eqref{inmodes}.

\section{Absorption cross section} \label{sec:absorption}
In this section, we show the procedure to compute the absorption cross section for arbitrary frequencies, which uses the solutions from the numerical integration scheme described in Sec.~\ref{sec:scalarfield}. We also show analytical approximate results in the low- and high-frequency regimes.

\subsection{Partial-waves approach}\label{sec:partial}
Using the standard partial wave method~\cite{Futterman:1988ni}, one can show that the total absorption cross section $\sigma$ of planar massless scalar waves impinging on a Schwarzschild BH surrounded by a spherical shell is given by
\be
\sigma=\sum_{l=0}^\infty \sigma_l,\label{eq:tabscs}
\ee
with
\be
\sigma_l=\frac{\pi}{\omega^2}(2l+1)\left|\frac{{\cal T}_{\omega l}}{{\cal A}_{\omega l}}\right|^2,\label{eq:pabscs}
\ee 
being the partial absorption cross section,\footnote{The partial absorption cross section, $\sigma_l$, corresponds to the absorption cross section for a fixed value of $l$.} where ${\cal T}_{\omega l}$ and ${\cal A}_{\omega l}$ are the coefficients appearing in Eq.~\eqref{inmodes}. This expression is algebraically the same as the one for a Schwarzschild BH without surrounding shells~\cite{Sanchez:1977si}. From Eq.~(\ref{eq:pabscs}), we see that the main ingredient to determine the absorption cross section for arbitrary frequencies is $\left|{\cal T}_{\omega l}/{\cal A}_{\omega l}\right|^2$, obtained using the procedure explained in Sec.~\ref{sec:scalarfield}.

\subsection{Low-frequency regime}\label{sec:low-frequency}

In order to determine the absorption cross section in the low-frequency regime, we start writing the solution of Eq.~(\ref{eq:de2}) for $\omega=0$ in two regions, namely, for $r<\rs$,
\be
U_{0l}^{-}(r)= \,C^{-}\, r\,P_l\left(y^{-}\right)+D^{-}\,r\, Q_l\left(y^{-}\right),\label{ineq}
\ee
and for $r>\rs$
\be
U_{0l}^{+}(r)=C^{+}\,r\, P_l\left(y^{+}\right)+D^{+}\,r\, Q_l\left(y^{+}\right),\label{outeq}
\ee
where $C^-,\,D^-,\,C^+,\,\text{and}\,D^+$ are constants, $P_l$ and $Q_l$ are the Legendre functions~\cite{abramowitz+stegun}. Also, we define
\be
y^{-}\equiv\frac{r}{\mh}-1,
\ee
and
\be
y^{+}\equiv\frac{r}{M}-1.
\ee
The regularity of the wave function at the BH horizon requires $D^-=0$. We can find relations among the remaining constants using the continuity of the scalar field at the shell position~(\ref{eq:continuity}) and the jump condition~(\ref{eq:jumpcond2})~(cf.~Appendix~\ref{app:coeffs}).

At the spatial infinity $r\rightarrow \infty$, the term proportional to $l(l+1)/r^2$ dominates the effective potential~(\ref{eq:effpotential}), so that we can write the solution for Eq.~(\ref{radialde}) as
\be
U_{\omega l}(x)\approx \omega x \left[(-i)^{l+1}\mathcal{A}_{\omega l}h{_l^{(1)*}}(\omega x)+(i)^{l+1}\mathcal{R}_{\omega l}h{_l^{(1)}}(\omega x)\right],\label{eq:hankelsolution}
\ee
where $h{_l^{(1)}}$ is the spherical Hankel function of first kind~\cite{abramowitz+stegun}.

If one considers the low-frequency regime $\omega x \ll 1$, it is possible to rewrite Eq.~\eqref{eq:hankelsolution} as
\be
U_{\omega l}\approx(-i)^{l+1}\mathcal{A}_{\omega l}\frac{(2)^{l+1} l!}{(2 l+1)!}(r \omega )^{l+1},\label{asympsolu}
\ee 
where we have considered $\omega r\approx \omega x$, for $r\rightarrow\infty$.

We can rewrite Eq.~\eqref{outeq} for $r\rightarrow \infty$, by using the following approximations:
\be
P_l\left(y^+\right)\approx\frac{2^{-l} (2 l)! \left(\frac{r}{M}\right)^l}{(l!)^2},\label{Pinfty}
\ee
and
\be
Q_l\left(y^+\right)\approx-i\frac{\pi  (2 l-1)\text{!!} \left(\frac{r}{M}\right)^l}{2 l!}.\label{Qinfty}
\ee

Inserting Eqs.~\eqref{Pinfty} and \eqref{Qinfty} in Eq.~\eqref{outeq}, we obtain an asymptotic expression which can be compared with Eq.~\eqref{asympsolu}, and we find an expression for the constant $C^-$~[cf.~Eq.~\eqref{CB}], which contains the dependence on the frequency.

In the regime $\omega x \ll 1$, the radial solution near the event horizon $r\approx r_H$ behaves as [see Eq.~\eqref{inmodes}]
\be
U_{\omega l}\approx \mathcal{T}_{\omega l}.\label{solhor}
\ee
Using Eqs.~\eqref{CA}--\eqref{CB} together with Eqs.~\eqref{ineq} and \eqref{outeq}, we obtain a complete solution in the low-frequency regime. By taking the limit $r\rightarrow r_H$ in Eq.~\eqref{ineq}, we can compare the resulting expression with Eq.~\eqref{solhor}, obtaining
\be
\frac{\mathcal{T}_{\omega l}}{\mathcal{A}_{\omega l}}\approx (-i)4\mh\omega + \mathcal{O}\l(\omega^{l+1}\r), \label{tracoefflf}
\ee 
 where the linear term in the frequency is related to the mode $l=0$.

 We have that, for $\omega\approx0$, combining Eq.~\eqref{tracoefflf} and Eq.~\eqref{eq:pabscs}
\be
\sigma_\text{lf} \approx \sigma_0\approx 16\pi\mh^2=4\pi r{_H^2},\label{eq:lfresult}
\ee
which stablishes that in the low-frequency regime only the mode $l=0$ contributes to the absorption cross section. Thus, the low-frequency regime of the absorption cross section, $\sigma_\text{lf}$, is given by the area of the event horizon, in agreement with the results for isolated spherically symmetric BHs~\cite{Das}.
\subsection{Null geodesics: high-frequency regime}\label{sec:high}

\begin{figure}
\includegraphics[width=\columnwidth]{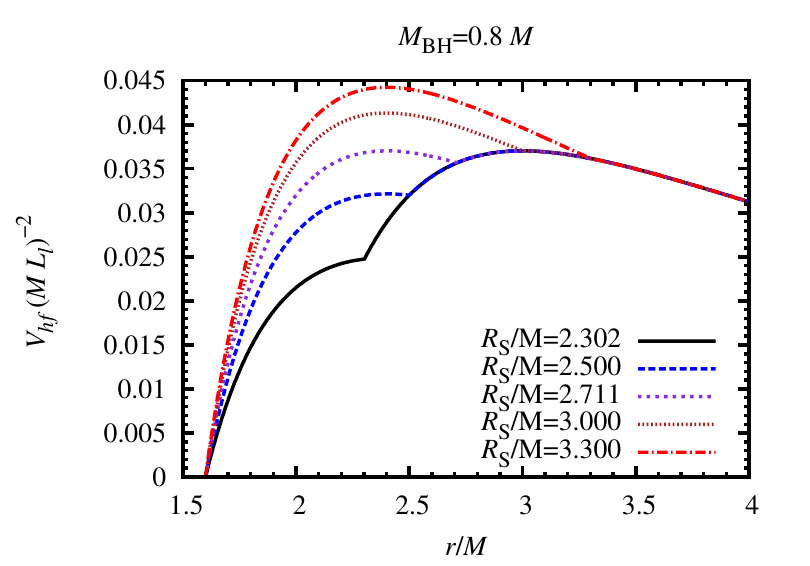}%
\caption{Effective potential $V_{hf}$, as a function of the radial coordinate for some values of the shell radius. Here, we assume $\mh=0.8 M$.}
\label{fig:potential_g}%
\end{figure}

In the high-frequency limit, our absorption/scattering problem can be reduced to the one of a stream of particles propagating along null geodesics. Without loss of generality, let us focus on null geodesics on the equatorial plane of the spherically symmetric spacetime under investigation. The Lagrangian associated to our problem is given by
\be
\mathcal{L}=\frac{1}{2}\l(-A\dot{t}^2+\frac{1}{B}\dot{r}^2+r^2\dot{\phi}\r)\equiv0,
\label{eq:lagragian}
\ee
where the overdot indicates a derivative with respect to the affine parameter of the curve. From the Lagrangian \eqref{eq:lagragian}, we have the following conserved quantities:
\bea
E=-\frac{\pa \mathcal{L}}{\pa \dot{t}}&=&A(r) \dot{t},\\
L=\frac{\pa \mathcal{L}}{\pa \dot{\phi}}&=&r^2\dot{\phi}.
\eea
Therefore, Eq.~\eqref{eq:lagragian} can be rewritten as
\be
\frac{A}{B}\dot{r}^2+V_{hf}(r)=E^2,
\label{eq:effective}
\ee
where 
\be
V_{hf}\equiv L^2 A/r^2,\label{eq:effpotgeo}
\ee
is the classical effective potential. When compared to an isolated Schwarzschild BH, the effective potential $V_{hf}$ has some interesting new features. From FIG.~\ref{fig:potential_g}, we can see that the potential derivative is discontinuous at the shell's position. Moreover, depending on the shell's location, we can have two local maxima, which indicates the existence of two unstable light rings (see discussion bellow).

Using Eq.~\eqref{eq:lagragian}, one can show that the radius of circular null geodesics $r_l$ and the ratio ${L_l}/{E_l}$ are given by~\footnote{The index $l$ denotes quantities related to the light ring.}
\bea
r_l =\frac{2A(r_l)}{A^{\prime}(r_l)},\label{eq:light}\\
b_l\equiv\frac{L_l}{E_l}=r_l\,A(r_l)^{-1/2},
\label{eq:bc}
\eea
respectively. For an isolated Schwarzschild BH, $b_l$ is the critical impact parameter for which a massless particle coming from the infinity ends up in an unstable circular orbit. For $b\equiv L/E<b_l$, the particle gets absorbed by the black hole, and for $b>b_l$, the particle is scattered to infinity. However, in the case of a DBH, as we shall see, the situation is not that simple. 

\begin{figure}
\includegraphics[width=\columnwidth]{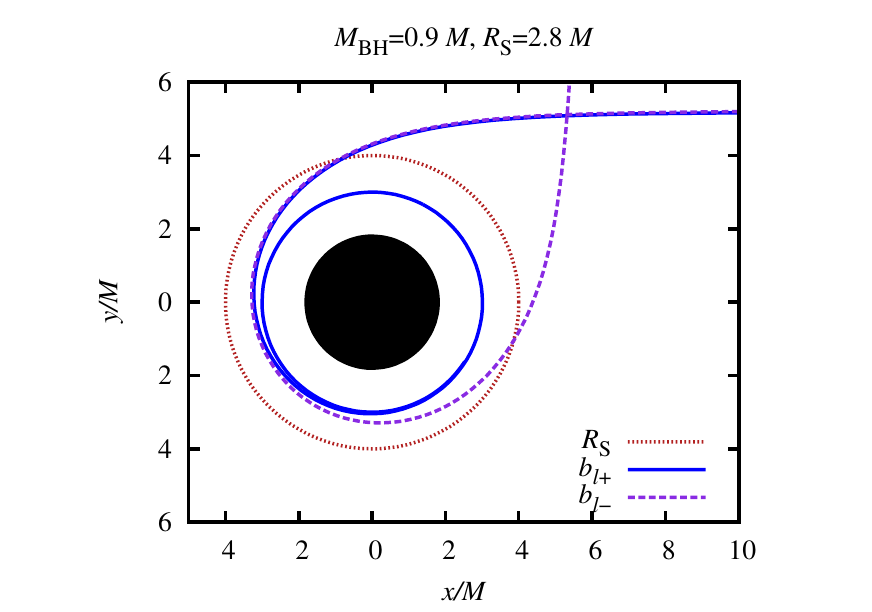}%
\caption{Null geodesics impinging from infinity with impact parameters $b_{l-}$ and $b_{l+}$. Here we have chosen~$\rs=2.8M$ and $\mh=0.9M$. In this case we have that the critical impact parameter corresponds to $b_{l+}$, since $b_{l+}<b_{l-}$, and the geodesic with the impact parameter $b_{l-}$ is scattered to infinity (dashed line). The dotted line represents the location of the shell at $\rs$.}
\label{fig:geo_bin_bex}%
\end{figure}

The presence of surrounding matter modifies the light ring structure. For the case of the DBH described in Sec.~\ref{sec:metric}, depending on the position of the spherical shell, we have three possibilities, namely,
\be
r_l=\l\{
\begin{array}{cl}
3M, &\text{ if } \rs< 3\mh,\\
3M \text{ and } 3\mh, &\text{ if } 3\mh <\rs< 3M,\\
3\mh, &\text{ if } \rs>3M.
\end{array}\r.
\label{eq:condgeo}
\ee 
Accordingly, we may have two different constants $b_l$ associated to unstable circular orbits. Since we are interested in absorption/scattering properties, we shall be concerned with the critical impact parameter. The critical impact parameter $b_c$ is such that for $b<b_c$ the light rays get captured by the BH. The only situation in which there can be ambiguities is the case $ 3\mh<\rs<3M$.\footnote{Examples of this situation are given by the plots for $\rs/M=2.500$~and~$2.711$, in FIG.~\ref{fig:potential_g}.} For this case, the critical impact parameter will be 
\be
b_c={\rm min}(b_{l+},b_{l-}), \label{eq:criticalimpactpar}
\ee
where
\be
b_{l+}=3\sqrt{3}M~ {\rm and }~ b_{l-}=3\sqrt{3}\mh/{\sqrt{\alpha}}
\label{eq:blplus}
\ee
are related to the light rings at $3M$ and $3\mh$, respectively. This situation is illustrated in FIG.~\ref{fig:geo_bin_bex}, where we have fixed the shell radius at $\rs=2.8M$ and the BH relative mass to $\mh=0.9M$. For this configuration one can show that $b_{l-}>b_{l+}$, so that we have $b_c=b_{l+}$.

Following the above analysis, the capture cross section, which corresponds to the high-frequency limit of the absorption cross section, is given by
\be
\sigma_\text{geo}= \pi b_c^2.\label{eq:capturecs}
\ee

\begin{figure}
\includegraphics[width=\columnwidth]{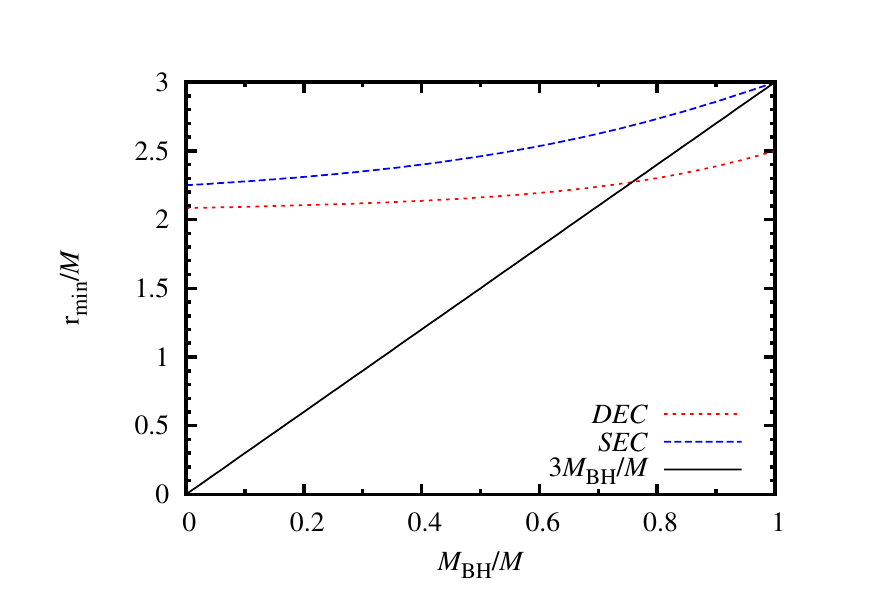}%
\caption{The minimum value for the shell position, $r_\text{min}$, according to SEC and DEC obtained through Eqs.~\eqref{eq:rmin_sec} and \eqref{eq:rmin_dec}, respectively, as a function of the mass ratio $\mh/M$. For SEC, the minimum value allowed for the shell position is always larger than $3\mh$. However, if one considers DEC, the shell can be placed inside~$3\mh$ for some values of $\mh/M$.}
\label{fig:shellposition}%
\end{figure}

We shall explore in more details the three possibilities described by Eq.~\eqref{eq:condgeo}. We can impose energy conditions to the shell, and this shall naturally restrict the shell position. In Appendix \ref{app:ec}, we show how the strong energy condition (SEC) and dominant energy condition (DEC) impose a minimum radius to the location of the shell. In FIG.~\ref{fig:shellposition}, we plot the lowest acceptable value for the shell position, $r_{\rm min}$, as a function of the BH mass, according to DEC and SEC, and compare with $3\mh$. Interestingly, DEC enables all three cases listed in Eq.~\eqref{eq:condgeo}, while for SEC we only have two possibilities (those for which $\rs>3\mh$). We note that $r_{\rm min}$ is always bigger than $2M$, despite the energy condition imposed, otherwise the whole configuration would naturally collapse into an isolated Schwarzschild BH with mass $M$. 

 We are interested in the effect of the spherical shell in the capture/scattering of null geodesics. From Eq.~\eqref{eq:effective}, we have that the geodesics can be described by
\be
\l(\frac{du}{d\phi}\r)^2-\frac{B}{b^2 A}+B u^2=0,
\label{eq:geodesic}
\ee
where we have defined $u\equiv r^{-1}$.

\begin{figure*}
\includegraphics[width=\columnwidth]{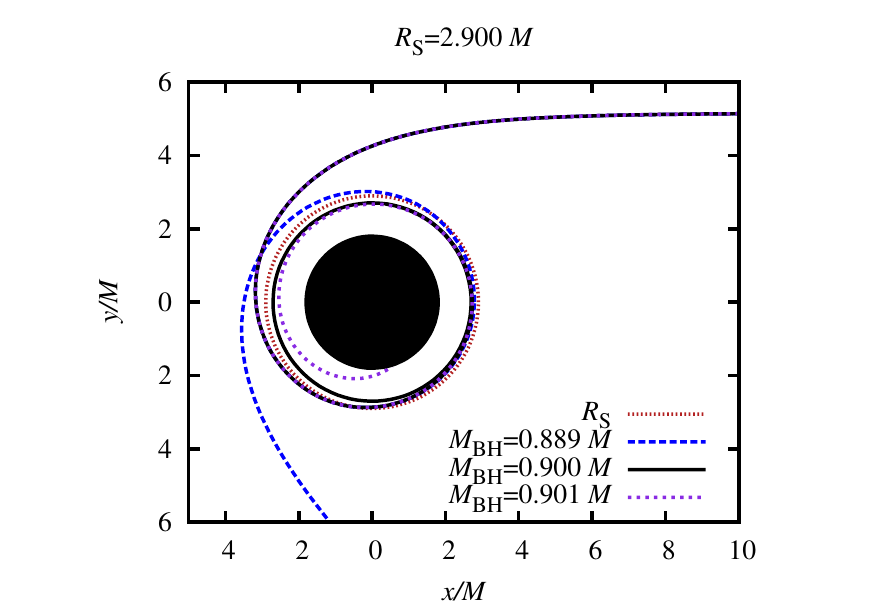}\includegraphics[width=\columnwidth]{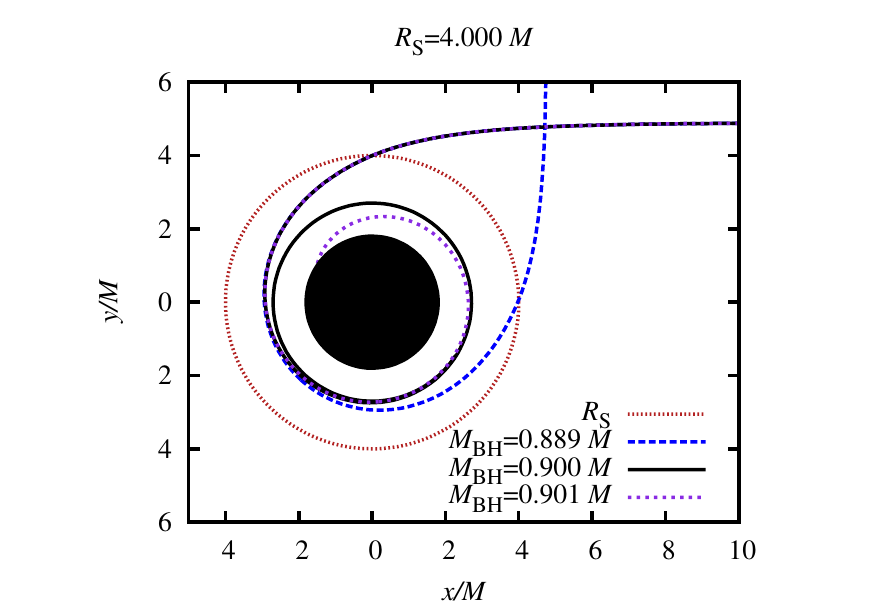}
\caption{Null geodesics coming from infinity with a fixed impact parameter. In the left and right panels we choose $\rs =2.900M$ and $\rs=4.000M$, respectively, and the impact parameter to be the critical one for $\mh=0.900M$. The dark region in both panels represents the BH with the largest event horizon.}%
\label{fig:geo}%
\end{figure*}

In order to illustrate the influence of the BH relative mass in the geodesic motion, we shall analyze in details three different geodesics with $\mh/M=0.899$, $0.900$, and $0.901$, for $\rs=2.900M$ and also for $\rs=4.000M$. The results are shown in FIG.~\ref{fig:geo}, where we exhibit cases for $3\mh<\rs<3M$~($\rs=2.900M$, left panel) and $\rs>3M$~($\rs=4.000M$, right panel). The impact parameter of the geodesics is chosen to be the critical one related to the case $\mh=0.9M$. For a smaller value of  $\mh/M$, the geodesic is scattered to infinity (dashed line). For a bigger mass ratio, the geodesic is captured by the BH (dotted line). Within this high-frequency analysis, we conclude that, for a fixed shell position~$\rs$, the larger the ratio $\mh/M$ is, the higher are the chances for a null geodesic to be absorbed.

\begin{figure*}
\includegraphics[width=\columnwidth]{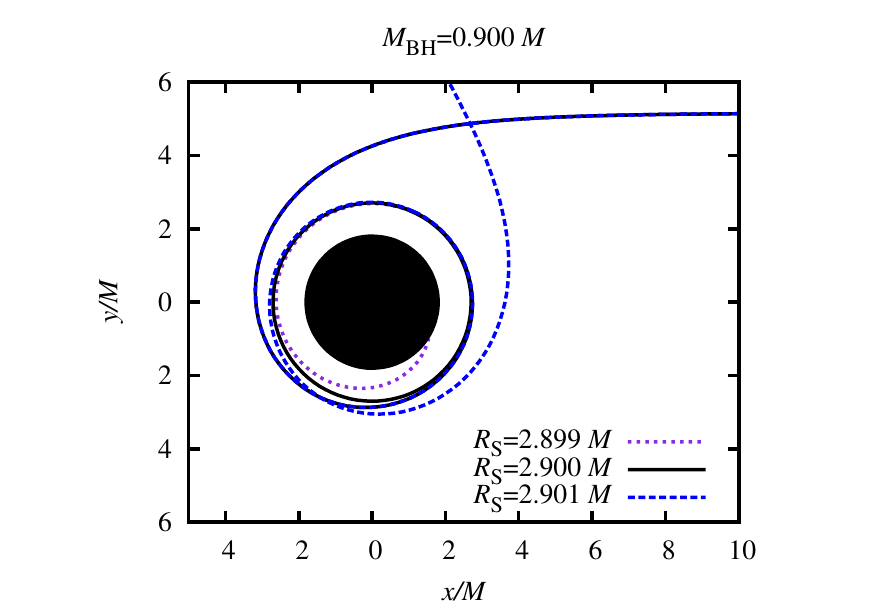}\includegraphics[width=\columnwidth]{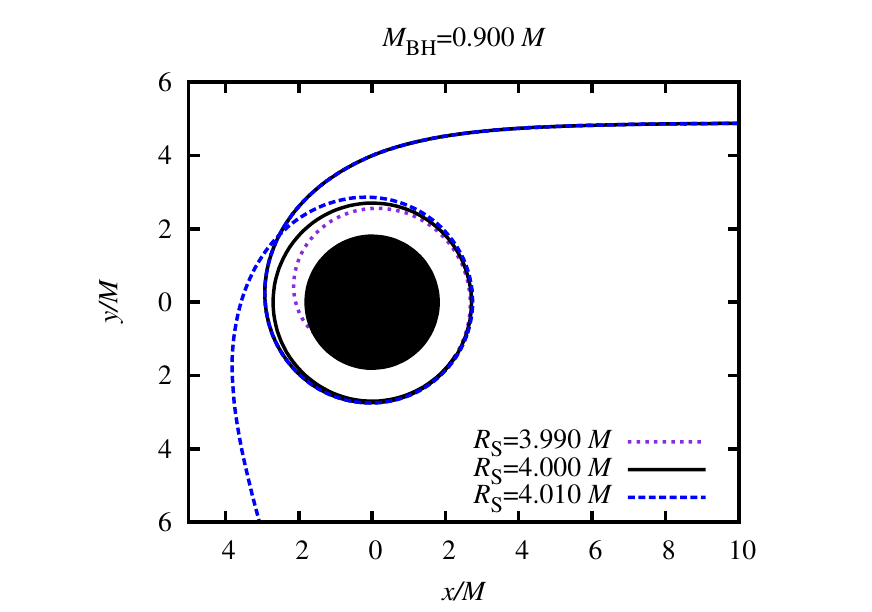}
\caption{As in FIG.~\ref{fig:geo}, we plot null geodesics impinging from infinity upon the DBH with a fixed impact parameter. Here we fix the ratio~$\mh/M$, and change the shell position. We choose $\mh =0.900M$ and $b$ to be the critical one for $\rs=2.900M$~(left panel) and $\rs=4.000M$~(right panel).}%
\label{fig:geo_2}%
\end{figure*}

The position of the shell is also important in the geodesic behavior. We survey its influence in the geodesic motion analyzing six different geodesics for $\mh=0.900M$, with $\rs/M=2.899$, $2.900$, $2.901$, $3.990$, $4.000$, and $4.010$. The results for these choices are shown in FIG.~\ref{fig:geo_2}. We have chosen the impact parameter to be the critical one related to the unstable circular geodesic (dark solid line) associated with the shell radius $\rs=2.900M$~(left panel), and $\rs=4.000M$~(right panel). For a smaller shell radius, the geodesic is captured by the BH (dotted line). For a bigger shell radius, the geodesic is scattered to infinity (dashed line). Therefore, for a fixed mass ratio~$\mh/M$, the smaller the shell radius is, more the configuration absorbs null geodesics.

\begin{figure}
\includegraphics[width=\columnwidth]{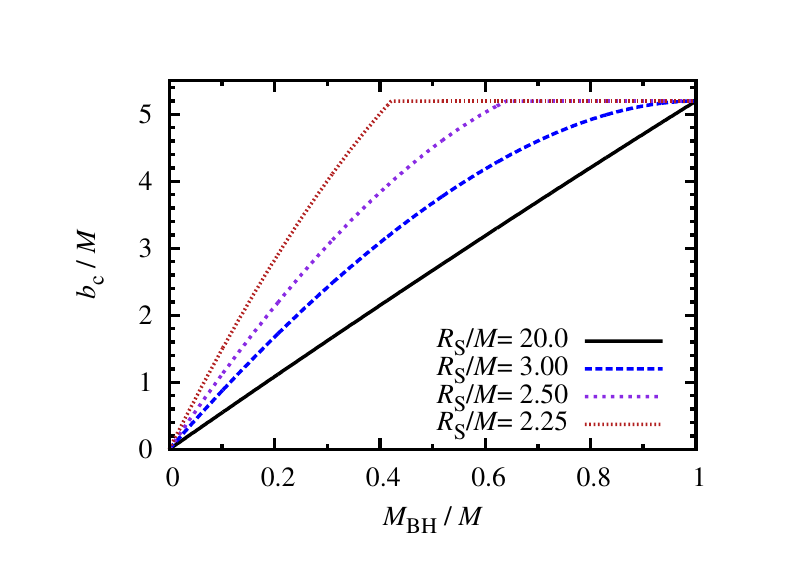}%
\caption{Critical impact parameter $b_c$ as a function of the BH relative mass $\mh/M$, for some values of the shell radius $\rs$.}
\label{fig:impact_par}%
\end{figure}

The above features can be confirmed by analyzing how the critical impact parameter varies with $\rs$ and $\mh/M$. In FIG.~\ref{fig:impact_par}, we plot the critical impact parameter, according to Eqs.~\eqref{eq:bc}--\eqref{eq:criticalimpactpar}, as a function of the BH relative mass, for some values of the shell radius. As we can see, for a fixed value of~$\rs/M$, the critical impact parameter increases as the relative mass of the BH increases, up to a point in which~$b_{l-}=b_{l+}$; then it equals~$b_{l+}=3 \sqrt{3}M$, regardless the value of the ratio~$\mh/M$. Moreover, for a fixed value of~$\mh/M$, the critical impact parameter increases as the radius of the shell decreases. 
When the mass of the BH is set to zero, we see that the critical impact parameter also goes to zero. This is expected, since there would be no absorption if there were no BH.

Another interesting feature arises when the thin spherical shell is far away from the BH. When this happens, from the null geodesics point of view, the whole system behaves similarly to a single BH with a mass $\mh$. This can be seen directly from the metric function $A(r)$, since
\be
A(r)\sim 1-\frac{2\mh}{r}+{\cal O}(\rs^{-1}),\label{eq:Aexpansion}
\ee
when $\rs\gg\mh$. This can be understood in terms of the surface energy density of the spherical shell, which decays quickly with its position~[cf. Eq.~\eqref{eq:density}]. To illustrate this, in FIG.~\ref{fig:impact_par}, we also show the case $\rs=20 M$, for which the critical impact parameter behaves almost as a straight line $b_c\sim 3 \sqrt{3}\mh$, resembling that of an isolated Schwarzschild BH~(with mass $\mh$).
	
\section{Results}\label{sec:results}
\begin{figure*}
\includegraphics[width=\columnwidth]{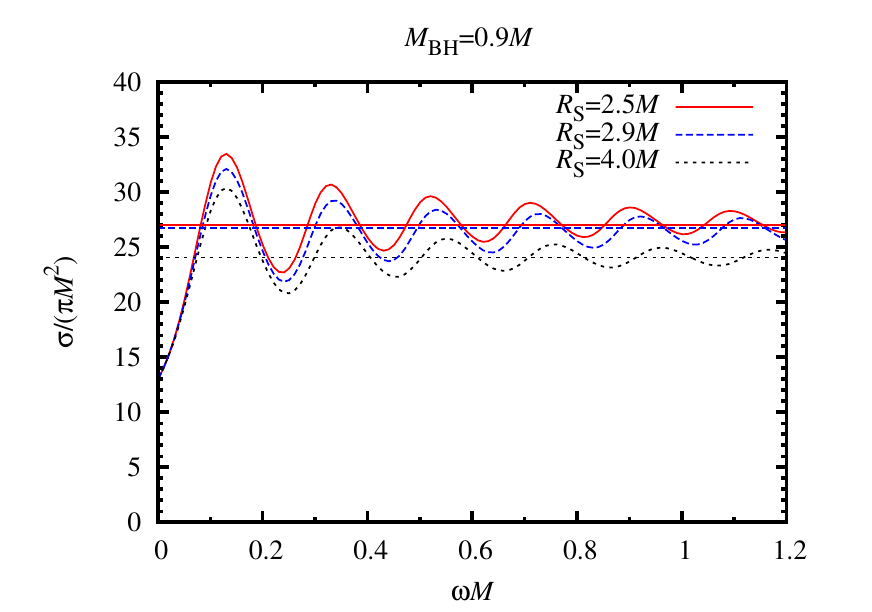}\includegraphics[width=\columnwidth]{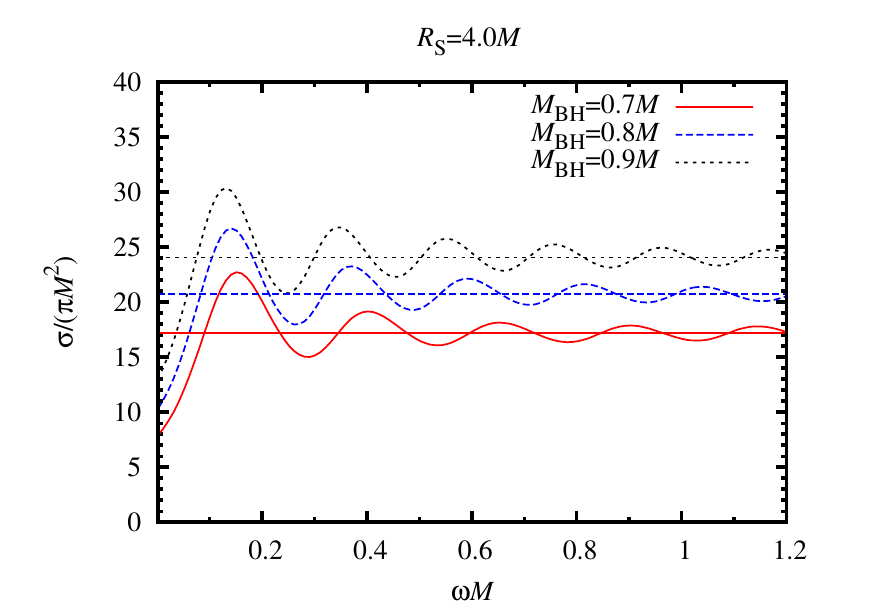}
\caption{LEFT: The total absorption cross section with $\mh=0.9M$ and $\rs/M=2.5,\,2.9,\,\text{and}\,\,4.0$. DEC is fulfilled for all three shell positions, while SEC does not allow the shell to be located at $\rs=2.5 M$ with $\mh=0.9M$, as it can be seen in FIG.~\ref{fig:shellposition}. RIGHT: The total absorption cross section for a fixed shell position $\rs=4.0M$ and different BH relative masses $\mh/M=0.7,\,0.8,\,\text{and}\,\,0.9$. In these cases, both DEC and SEC are satisfied. The horizontal lines correspond to the high-frequency limit in each case.}
\label{DBH_MH0p9}
\end{figure*}

\begin{figure*}
\includegraphics[width=\columnwidth]{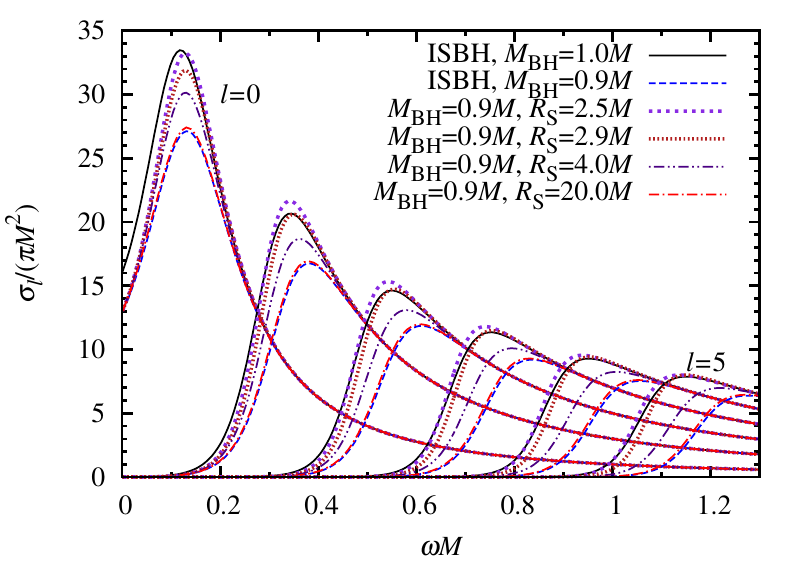}\includegraphics[width=\columnwidth]{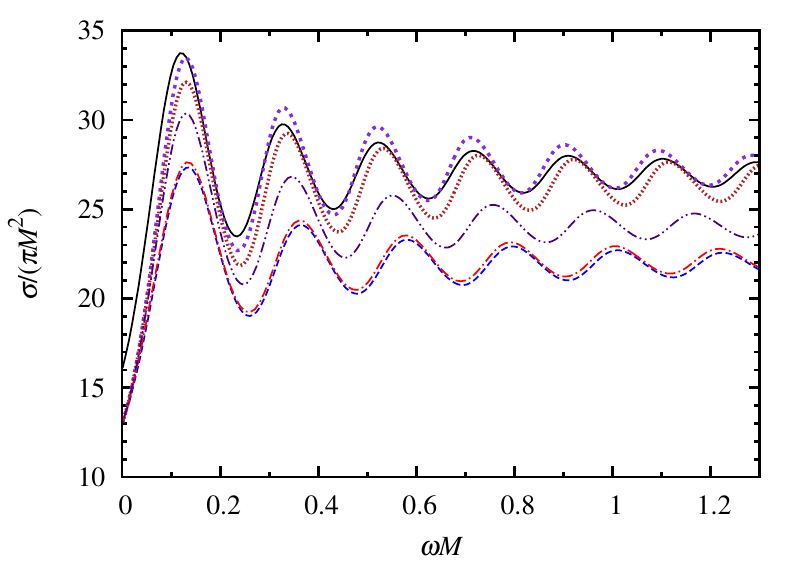}
\caption{LEFT: Comparison between the partial absorption cross sections, $\sigma_l$, for DBHs and ISBHs, with $l=0,\,1,\,2,\,3,\,4$ and $5$. RIGHT: The total absorption cross sections for DBHs compared with the ones of ISBHs.}
\label{DBH_Schwarzschild}
\end{figure*}

In this section, we present a selection of our numerical results for the scalar absorption of a Schwarzschild BH surrounded by a thin spherical shell of matter. As a general behavior for mid-to-high frequencies, we note that the total absorption cross section, similarly to the case of an isolated Schwarzschild BH~(ISBH), presents a regular oscillatory pattern around its high-frequency limit.

In the left panel of FIG.~\ref{DBH_MH0p9}, we show the total absorption cross section for a fixed black hole relative mass $\mh=0.9M$ and different positions of the shell~$\rs/M=2.5,\,2.9,\,\text{and}\,\,4.0$. As the frequency increases, we note that the total absorption cross section increases from the area of the BH event horizon, according to Eq.~\eqref{eq:lfresult}, and then oscillates regularly around the high-frequency limit given by the capture cross section of null geodesics, according to Eq.~\eqref{eq:capturecs}, which is represented by a horizontal line. When we keep the value of $\mh/M$ fixed, we see that the absorption decreases as the shell is positioned further away from the BH. Furthermore, in the right panel of FIG.~\ref{DBH_MH0p9}, where we consider the value of the shell position fixed at~$\rs=4.0M$ and different BH relative masses $\mh/M=0.7,\,0.8,\,\text{and}\,\,0.9$, we note that the absorption increases as the mass ratio~$\mh/M$ increases.

In FIG.~\ref{DBH_Schwarzschild}, we show the results for the partial~($\sigma_l$, left panel) and total~($\sigma$, right panel) absorption cross sections for Schwarzschild BHs surrounded by a thin spherical shell of matter and compare with the results for ISBHs. We see that when the shell is located close to the BH, the values of the cross sections for DBHs are similar to the ones for an ISBH with mass $M$. This is in agreement with the fact that, when the shell is near to the BH, the whole system behaves similarly to an isolated BH with mass $M$. Also, we note that for large values of the shell radius~$\rs$, the values of the absorption cross sections for DBHs become closer to the results for an ISBH with mass $\mh$, in accordance with~Eq.~(\ref{eq:Aexpansion}).

\section{Final remarks}\label{sec:remarks}
We have considered planar massless scalar waves impinging upon a Schwarzschild BH surrounded by a thin spherical shell and determined its absorption spectrum. Our numerical results are in full agreement with the analytical formulas obtained in the low- and high-frequency approximations. In particular, for the low-frequency regime, we have confirmed with our numerical results that the total absorption cross section is given by the area of the BH event horizon.  

We have shown that for the same $\mh/M$ ratio the absorption is bigger when the shell is closer to the BH. Moreover, for a fixed shell position, the absorption cross section increases as the BH relative mass increases. Also, when we compare the absorption of a DBH with an ISBH, we have shown that, as the shell is placed further away from the BH the absorption cross section of the DBH tends to the result of an ISBH with the same mass ratio $\mh/M$ of the BH inside the shell. This is in accordance with the fact that as the shell is positioned further away from the BH the presence of the shell becomes less relevant to the absorption process. On the other hand, as we approximate the shell to the BH, we note that the absorption becomes similar to the case of an isolated BH with mass $M$ equal to the whole system (BH with shell) mass.

The results presented in this paper generically show that the presence of matter around black holes can modify their absorption properties. The modifications due to the surrounding matter intensify for higher densities of the shell. Moreover, the position of the shell also plays a important role in the absorption properties.

Another interesting feature that arises due to the matter surrounding the black hole is the appearance of a second light ring. We have seen that the two light rings happen for certain values of the mass and position of the thin shell. We expect that other matter configurations would also present two light rings, for instance, when considering thick shells. Between the two light ring positions, there is a local minimum of the potential, similar to the one that appears in the case of gravastars~\cite{Mazur:2004fk}. In the thick shell case, this minimum point can be related to the existence of a \textit{stable} light ring. Moreover, stable light rings can support long-lived modes in the eikonal limit~\cite{Cardoso:2014sna}. These long-lived modes can source nonlinear instabilities, and therefore any kind of exotic matter that allows the existence of such feature may eventually collapse to the BH, putting further constraints on the matter surrounding BHs.

\begin{acknowledgments}
The authors would like to thank Conselho Nacional de Desenvolvimento Cient\'ifico e Tecnol\'ogico (CNPq), Coordena\c{c}\~ao de Aperfei\c{c}oamento de Pessoal de N\'ivel Superior (CAPES), Funda\c{c}\~ao Amaz\^onia de Amparo a Estudos e Pesquisas do Par\'a (FAPESPA), and International Research Staff Exchange Scheme (IRSES) for partial financial support.
\end{acknowledgments}
%

\appendix
\begin{widetext}
\section{Coefficients of Eqs.~\eqref{ineq} and \eqref{outeq}}\label{app:coeffs}

The coefficients appearing in Eq.~\eqref{outeq} can be rewritten in terms of the coefficient $C^-$, as follows:
\begin{eqnarray}
C^{+} & = & C^{-} \left[\frac{-P_{l}\left(\frac{R_{S}}{\mh}-1\right)\left( \eta+l(\alpha-1)Q_{l}\left(\frac{R_{S}}{M}-1\right)R_{S}\right) + \beta Q_{l}\left(\frac{R_{S}}{M}-1\right)}{\tau}\right]\label{CA},
\end{eqnarray}
and
\be
D^{+} = C^{-}\left[\frac{P_{l}\left(\frac{R_{S}}{\mh}-1\right)\left(\chi+l(\alpha-1)P_{l}\left(\frac{R_{S}}{M}-1\right)R_{S}\right)- \beta P_{l}\left(\frac{R_{S}}{M}-1\right)}{\tau}\right] \label{DA},
\ee
where
\be
\eta=M \left[\left(\sqrt{\alpha}+l\sqrt{\alpha}+l-1\right)Q_{l}\left(\frac{R_{S}}{M}-1\right)+\left(\sqrt{\alpha}+1\right)(l+1)Q_{l+1}\left(\frac{R_{S}}{M}-1\right)\right],
\ee
\be
\chi=M \left[\left(\sqrt{\alpha}+l\sqrt{\alpha}+l-1\right)P_{l}\left(\frac{R_{S}}{M}-1\right)+\left(\sqrt{\alpha}+1\right)(l+1)P_{l+1}\left(\frac{R_{S}}{M}-1\right)\right],
\ee
\be
\beta=\sqrt{\alpha}\left[\left(-\sqrt{\alpha}+l\sqrt{\alpha}+l+1\right)P_{l}\left(\frac{R_{S}}{\mh}-1\right)+\left(\sqrt{\alpha}+1\right)(l+1)P_{l+1}\left(\frac{R_{S}}{\mh}-1\right)\right],
\ee
and
\be
\tau=\left(\sqrt{\alpha}+1\right)M(l+1)\left[P_{l+1}\left(\frac{R_{S}}{M}-1\right)Q_{l}\left(\frac{R_{S}}{M}-1\right)-P_{l}\left(\frac{R_{S}}{M}-1\right)Q_{l+1}\left(\frac{R_{S}}{M}-1\right)\right].
\ee

The comparison between Eqs.~\eqref{asympsolu} and \eqref{outeq}, rewritten with the aid of Eqs.~\eqref{Pinfty}~and~\eqref{Qinfty}, gives the following result for the coefficient $C^-$,

\begin{eqnarray}
C^{-} = \frac{\mathcal{A}_{\omega l}4^{l+1}\omega(l!)^{3}(-iM\omega)^{l}\tau}{(2l+1)! \left[\epsilon-\beta \left(\pi2^{l}l!(2l-1)\text{!!}P_{l}\left(\frac{R_{S}}{M}-1\right)-2i(2l)!Q_{l}\left(\frac{R_{S}}{M}-1\right)\right)\right]}, \label{CB}
\end{eqnarray}
where
\be
\epsilon= P_{l}\left(\frac{R_{S}}{\mh}-1\right)\left[\pi2^{l}l!(2l-1)\text{!!}\left(\chi+l(\alpha-1)R_{S}P_{l}\left(\frac{R_{S}}{M}-1\right)\right)-2i(2l)!\left(\eta+l(\alpha-1)R_{S}Q_{l}\left(\frac{R_{S}}{M}-1\right)\right)\right].
\ee

\end{widetext}

\section{Energy conditions}\label{app:ec}

Using the line element (\ref{eq:metric}), it can be shown that the non-zero components of the Einstein tensor $G_{a}^{\,\,b}$~\cite{wald} are
\begin{eqnarray}
G_{t}^{\,\,t} & = &{ \frac{B'}{r}+\frac{B}{r^2}-\frac{1}{r^2
   }},\label{G00}\\
G_{r}^{\,\,r} & = &{ \frac{B A'}{r
   A}+\frac{B}{r^2}-\frac{1}{r^2}},\label{G11}\\
G_{\theta}^{\,\,\theta} & = & {\frac{B A''}{2 A}+\frac{A'
   B'}{4 A}-\frac{B A'^2}{4
   A^2}+\frac{B A'}{2 r
   A}+\frac{B'}{2 r}},\label{G22}\\
G_{\phi}^{\,\,\phi} & = &  G_{\theta}^{\,\,\theta}.\label{G33}
\end{eqnarray}

Recalling Einstein's equations, $G_{a}^{\,\,b}=8\pi T_{a}^{\,\,b}$, we note that, from Eqs.~\eqref{G00}-\eqref{G33}, the associated energy-momentum tensor $T_{a}^{\,\,b}$ has only nonzero components for $a=b$. Therefore, the energy density is given by $\rho = -T_{t}^{\,\,t}$, and the pressures along a direction $j$ are given by $p_j = T_{j}^{\,\,j}$. Here, the index $j$ refers to angular coordinates $\theta$ and $\phi$, with no implicit summation involved. 

Imposing energy conditions, namely SEC and DEC, to the content of the massive shell surrounding a Schwarzschild BH, which is composed by perfect fluid, we can obtain a lower limit for the position of the shell, as follows.

\subsection{Strong energy condition (SEC)}\label{app:strong}
Let us first examine the restriction to the position of the shell following from the imposition of the SEC. Introducing the quantity
\begin{equation}
U_{a}^{\,\,b}=\lim_{\epsilon\rightarrow0^{+}}\int_{R_{s}-\epsilon}^{R_{s}+\epsilon}\, T_{a}^{\,\,b}{\sqrt{g_{rr}}}dr,\label{integration}
\end{equation}
it can be shown that the following relation holds,
\begin{equation}
{U_{{t}}^{\,\,{t}} + U_{{j}}^{\,\,{j}}}\geq 0,\label{eq:SEC1}
\end{equation}
with no implicit sum on $j$.

Thus, substituting Eq.~\eqref{G00} into Eq.~\eqref{integration}, we are left with
\be
U_{t}^{\,\,t}={-}\frac{1}{8\pi}\lim_{\epsilon\rightarrow0^{+}}\int_{\rs - \epsilon}^{\rs+\epsilon}\,{\frac{1}{\sqrt{B}}}\left(\frac{B^{\prime}}{r}\right)dr.
\ee
Using the Einstein's equations and inserting Eq.~(\ref{G00}) in Eq.~(\ref{integration}), we obtain
\be
U_{t}^{\,\,t}=-\frac{1}{4\pi \rs}\left(\sqrt{1-\frac{2M}{\rs}}-\sqrt{1-\frac{2\mh}{\rs}}\right)\label{density}.
\ee

Analogously, we find that
\begin{equation}
U_{{\theta}}^{\,\,{\theta}}={\frac{1}{8\pi \rs}\left(\frac{1-\frac{M}{\rs}}{\sqrt{1-\frac{2M}{\rs}}}-\frac{1-\frac{\mh}{\rs}}{\sqrt{1-\frac{2\mh}{\rs}}}\right)}.\label{pressure}
\end{equation}

Now, using Eq.~(\ref{eq:SEC1}), with the aid of Eqs.~(\ref{density}) and (\ref{pressure}), we are left with 
\bea
r_\text{min}^{SEC}&=&\frac{3}{8}\left(\sqrt{9M^2-14M\mh+9\mh^2}\right.\nonumber\\
&+&\left.3M+3\mh\right).\label{eq:rmin_sec}
\eea
We note that when $\mh=M$, which corresponds to an ISBH, the minimum value for the shell position is $r_\text{min}^{SEC}=3\mh$. From~\eqref{eq:rmin_sec}, as shown in FIG.~\ref{fig:shellposition}, if one assumes SEC, the shell is always placed outside of $3\mh$.

\subsection{Dominant energy condition (DEC)}\label{app:dominant}
We may carry out a similar analysis considering DEC. By doing that, we obtain
\begin{equation}
{U_{{t}}^{\,\,{t}}}-\l|U_{{j}}^{\,\,{j}}\r|\geq0,\label{eq:DEC}
\end{equation}
for each $j=\theta,\,\phi$~(with no implicit summation involved).

Using Eq.~(\ref{eq:DEC}) together with Eqs.~(\ref{density}) and (\ref{pressure}),
it is possible to show that the lowest value of $\rs$, according to DEC, is given
by
\bea
r_\text{min}^{DEC}&=&\frac{5}{24}\left(\sqrt{25M^2-46M\mh+25\mh}\right.\nonumber\\
&+&\left.5M+5\mh\right).\label{eq:rmin_dec}
\eea
Considering $\mh=M$ in Eq.~\eqref{eq:rmin_dec}, the minimum value for the shell position is $r_\text{min}^{DEC}=5\mh/2$. Assuming DEC, the minimum value for the shell position is always smaller than the corresponding one related to SEC, for the same value of the ratio~$\mh/M$~(cf.~FIG.~\ref{fig:shellposition}).

\bibliography{refs}
\end{document}